\documentclass[sigconf]{acmart}

\AtBeginDocument{%
  }

\setcopyright{acmlicensed}
\copyrightyear{2026}
\acmYear{2026}
\acmDOI{10.1145/3744916.3787828}
\acmConference[ICSE]{IEEE/ACM 48th International Conference on Software Engineering}{April 12-18,
  2026}{Rio de Janeiro, Brazil}
\acmISBN{XXXX}

\acmSubmissionID{icse2026-p5525}

\usepackage{rotating}
\usepackage{enumitem}
\usepackage{bbding}

\newcommand{\hl}[1]{{#1}}
\newcommand{\name}[0]{{\textsf{DyMA-Fuzz}}}
\newcommand{\dice}[0]{{\textsf{DICE}}}

\newcommand{\multifuzz}[0]{{\textsf{MultiFuzz}}}
\newcommand{\semu}[0]{{\textsf{SEmu}}}
\newcommand{\icicle}[0]{{\textsf{Icicle}}}

\usepackage{makecell}
\usepackage{color}
\usepackage{multirow}
\usepackage{xspace}
\usepackage{float}
\usepackage{colortbl}

\newcommand{\targets}{94}

\usepackage{pifont}
\usepackage{xcolor}

\usepackage{tabularray}
\usepackage[framemethod=TikZ]{mdframed}
\usepackage{hhline}
\usepackage{circledsteps}

\pgfkeys{/csteps/inner ysep=3pt}
\pgfkeys{/csteps/inner xsep=4pt}
\pgfkeys{/csteps/inner color=white}
\pgfkeys{/csteps/outer color=black}
\pgfkeys{/csteps/fill color=black}

\begin{document}

\title{\textsc{DyMA-Fuzz}: Dynamic Direct Memory Access Abstraction for Re-hosted Monolithic Firmware Fuzzing}
\author{Guy Farrelly}
\affiliation{%
  \institution{Adelaide University}
  \city{Adelaide}
  \country{Australia}
}
\email{guy.farrelly@adelaide.edu.au}

\author{Michael Chesser}
\affiliation{%
  \institution{Adelaide University}
  \city{Adelaide}
  \country{Australia}
}
\email{michael.chesser@adelaide.edu.au}

\author{Seyit Camtepe}
\affiliation{%
  \institution{CSIRO Data61}
  \city{Sydney}
  \country{Australia}}
\email{seyit.camtepe@data61.csiro.au}

\author{Damith C. Ranasinghe}
\affiliation{%
  \institution{Adelaide University}
  \city{Adelaide}
  \country{Australia}}
\email{damith.ranasinghe@adelaide.edu.au}

\begin{abstract}
The rise of smart devices in critical domains---including automotive, medical, industrial---demands robust firmware testing. Fuzzing firmware in re-hosted environments is a promising method for automated testing at scale, but remains difficult due to the tight coupling of code with a microcontroller's peripherals. Existing fuzzing frameworks primarily address input challenges in providing inputs for Memory-Mapped I/O or interrupts, but largely overlook Direct Memory Access (DMA), a key high-throughput interface used that bypasses the CPU.

We introduce \name{} to extend recent advances in stream-based fuzz input injection to DMA-driven interfaces in re-hosted environments. It tackles key challenges---vendor-specific descriptors, heterogeneous DMA designs, and varying descriptor locations---using runtime analysis techniques to infer DMA memory access patterns and automatically inject fuzzing data into target buffers, without manual configuration or datasheets.

Evaluated on \targets{} firmware samples and 8 DMA-guarded CVE benchmarks, \name{} reveals vulnerabilities and execution paths missed by state-of-the-art tools and achieves up to 122\% higher code coverage.
These results highlight DyMA-Fuzz as a practical and effective advancement in automated firmware testing and a scalable solution for fuzzing complex embedded systems.
 
\end{abstract}



\maketitle

\section{Introduction}

Rapid proliferation of smart devices in safety-critical domains such as automotive, medical, and industrial control systems demands robust testing of embedded, monolithic firmware. These monolithic binaries interact closely with hardware, where malformed or malicious input can lead to hazardous real-world consequences~\cite{Goodin2017, Goodin2018}.

\begin{figure}[b]
    \centering
    \includegraphics[width=1.0\linewidth]{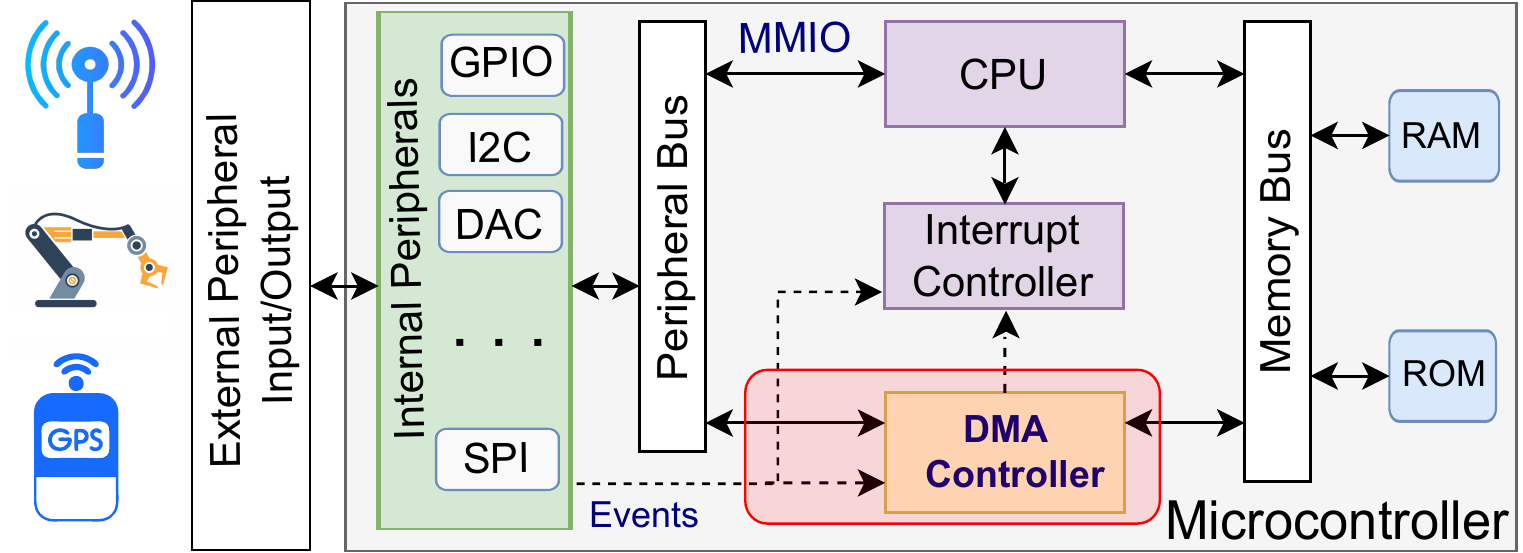}
    \caption{Simplified overview of a MCU. Data-flow is indicated with solid lines, event triggers such as interrupts are denoted with dashed lines. External inputs are delivered through peripherals. Many existing emulation-based fuzzing frameworks neglect the interactions highlighted in \color{red}red \color{black} due to the complexity of emulating DMA interactions.}
    \label{fig:mcu_overview}
\end{figure}

Fuzz testing firmware in re-hosted environments is proving an effective, efficient, and scalable method for automatically uncovering vulnerabilities. Re-hosting is unconstrained by the resource limits of the target microcontroller hardware and facilitates binary-only testing. However, application to firmware remains challenging due to the tight coupling with microcontroller resources and peripherals. As illustrated in Figure~\ref{fig:mcu_overview}, microcontroller units (MCUs) use a range of internal peripherals to communicate with an endless array of external input/output peripherals. Firmware interacts with peripherals using a mix of memory-mapped I/O (MMIO) registers, interrupts, and Direct Memory Access (DMA).
In general, facilitating fuzz inputs to firmware accessing a plethora of possible internal or external peripherals from many potential device manufacturers in an emulation environment is nontrivial.

Earlier fuzzers made progress towards handling firmware interactions with MMIO and interrupts~\cite{emberio2023,fuzzware2022,icicle2023,p2im2020,uemu2021,multifuzz2024,hoedur2023,splits2023}. However, DMA---a crucial input path used by high-throughput peripherals---remains a key challenge. Notably, prior work~\cite{dice2021} identified that more than 94\% of 32-bit microcontroller products contain DMA capabilities, and more than 25\% of analysed firmware samples from GitHub contain DMA related symbols. DMA operates independently of the CPU, transferring data between memory and peripherals as illustrated in Figure~\ref{fig:mcu_overview}. Transfers are configured via a transfer descriptor, a structure containing critical information such as a \textit{destination address}, \textit{source address} and \textit{buffer size} following a \textit{vendor-specific} DMA interface. The format of these descriptors is unknown to a re-hosted environment, but fuzzing DMA-driven code requires accurate identification and emulation of the described transfers. Importantly, unlike fuzz testing data injections to MMIO---where drivers often tolerate fuzzer-based manipulation of bits that should not be mutated---false positives in DMA emulation can corrupt memory directly, completely invalidating further execution.

\subsection{Challenges}
We elaborate on the unique challenges specific to fuzzing DMA enabled peripherals below (and provide a detailed study in Section~\ref{sec:dma-study}):
\begin{itemize}[itemsep=1pt,parsep=0pt,topsep=3pt,labelindent=0pt,leftmargin=7mm]
    \item[\Circled{C1}]\textbf{Lack of standards.~}Information in the DMA transfer descriptor used to configure DMA are non-standard---the internal layout and related variables vary by manufacturer or model.

    \item[\Circled{C2}]\textbf{Implicitly defined configuration values.~}Not all information required to infer the source, destination, and size of a transfer is always explicitly defined in the descriptor. Instead, some of the values can be statically defined by the manufacturer, applying to all transfers using that peripheral.

    \item[\Circled{C3}]\textbf{Varied descriptor locations.~}Not all DMA transfer descriptors are located in a predesignated MMIO region; some are stored in RAM, at locations designated by the firmware.

    \item[\Circled{C4}]\textbf{Ever growing number of interfaces.~}New MCU releases can introduce new, unique transfer descriptor layouts.

    \item[\Circled{C5}]\textbf{Multiple DMA descriptor layouts on a single MCU.~}Different, unique descriptor layouts can exist on one MCU, and information inferred from one descriptor layout may not apply to descriptors for another DMA-enabled peripheral.
    
\end{itemize}

To address the challenges, DICE~\cite{dice2021} uses a heuristic extension to a fuzzer capable of injecting data into MMIO peripherals, to attempt detection and injection of data into DMA buffers, but relies on strict assumptions about a descriptor's location and internal layout. This limits generalization across microcontrollers. Notably, many MCUs, as shown in our study in Section~\ref{sec:dma-study}, use multiple DMA controllers of varying complexity. In contrast, \semu{}~\cite{semu2022} uses datasheets as a ground truth to automatically describe a DMA controller's implementation to achieve higher fidelity. 
The approach relies on the availability of complete and accurate external information, and human effort is still required when the correct outcomes cannot be identified from parsing the provided information---this is particularly common for DMA~\cite{perry2024}. This manual effort, and incomplete or inconsistent documentation also limits scalability.

Our efforts aim to develop a \textit{versatile} yet accurate automated approach to dynamically identify DMA descriptors, without reliance on external information. Using insights from a study of DMA controllers from leading MCU manufacturers, we simplify and scale descriptor inference to fuzz a wider range of MCU families.

\subsection{Contributions}
To address the challenge of fuzzing monolithic firmware relying on DMA transfers in emulation, we make the following contributions:
\begin{itemize}[itemsep=2pt,parsep=0pt,topsep=3pt,labelindent=0pt,leftmargin=5mm]
    \item Develop analysis techniques for accurate, dynamic identification of key information from diverse DMA transfer descriptors by conducting a systematic study and analysis of DMA interfaces from leading microcontroller manufacturers.

    \item Build \name{}, a prototype implementing our automated dynamic analysis techniques. \name{} avoids dependence on external datasheets or pre-defined interfaces and integrates with a state-of-the-art input generation method to create dedicated stream-based inputs for DMA channels. We open-source our code at \url{https://github.com/DyMA-Fuzz/DyMA-Fuzz}.
    
    \item In a systematic evaluation with a dataset of \targets{} binaries (including a CVE benchmark), we demonstrate our approach allows \name{} to significantly outperform state-of-the-art DMA fuzzers in code exploration and bug discovery.
    
\end{itemize}

\begin{figure*}[t!]
    \centering
    \includegraphics[width=1.0\linewidth]{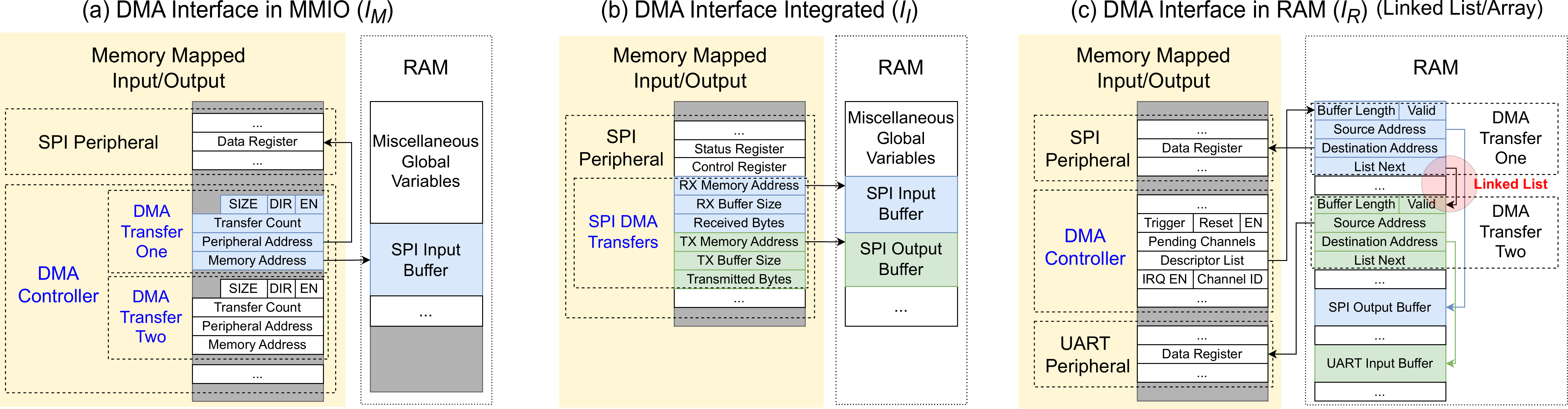}
    \caption{(a)~Example MMIO layout of a Multi-Channel DMA Controller (Based on STM32F0). A reserved region of memory is used to configure DMA transfers for all peripherals. Several transfers can be configured within a single controller, utilizing separate channels. (b)~Example MMIO layout of an Integrated DMA interface within an SPI peripheral (Based on nRF51). DMA transfers for each peripheral are configured within the peripheral's own MMIO space. (c)~Example MMIO and RAM layout of a list-based DMA interface (Based on SAM L10, the CC2538 uses an array of similar format, instead of a linked list).}
    \label{fig:listbased}
    \label{fig:integrated}
    \label{fig:multichannel}
\end{figure*}

\section{Re-hosting for Fuzzing Primer}
\noindent\textbf{Firmware Fuzzing.~}
Fuzzing uses a series of randomly mutated inputs to explore different functionality and identify memory corruption bugs in software. Fuzzing facilitates automated testing of software to support software engineers and security researchers discover bugs and vulnerabilities. While fuzzing is prevalent in the desktop application space, firmware fuzzing faces unique challenges due to the close coupling between firmware and hardware, and limited resources~\cite{wycinwyc2018}. Fuzzing of monolithic firmware has been approached from several angles, with varying levels of dependence on hardware. One method is to obtain the physical target platform, and use dedicated debugging interfaces to probe or modify the system state while the firmware is running~\cite{gdbfuzz2023,uafl2022,avatar2014,avatar2018,conware2021,pretender2019,inception2018}. However, limited resources on the target device, and limited debugger interface speeds restrict fuzzing throughput. Some frameworks function on original hardware, but require compilation from source to include a fuzzing interface to reduce or entirely remove dependence on hardware debuggers~\cite{shift2024}. However, these approaches still face scalability limits, as each parallel instance of tests requires an additional hardware device. Moving to a high-performance server or desktop PC allows scaling to tens, or hundreds of parallel instances.

\vspace{1mm}

\noindent\textbf{Firmware Re-hosting.~}
Re-hosting is the process of executing a non-native program in a desired environment, without modifying the program itself. This is challenging, but desirable for firmware, as the capabilities for debugging and analysis are limited by the microcontroller's resources~\cite{wycinwyc2018}. Achieving successful firmware re-hosting is naturally difficult; inputs and outputs are facilitated through peripherals without a standardised implementation. Successful re-hosting is dependent on achieving emulation accurate enough to pass any validation checks performed by the driver.

Some holistic re-hosting approaches attempt to emulate all peripherals on a microcontroller based on analysis of manufacturer provided information~\cite{perry2024,semu2022}. By generating precise rule-sets from this data, high fidelity emulation can be achieved. However, this can generate thousands of rules, increasing complexity. The generation of these rules is not always accurate, and manual analysis to identify and correct inaccuracies is nontrivial.

Contrastingly, several other methods approach re-hosting from a simplified rule-set, and without any dependence on manufacturer supplied information. Several works focused on Memory Mapped Input Output (MMIO) apply one of only a few models to each register~\cite{p2im2020,uemu2021,fuzzware2022}, and some function with just a single model~\cite{emberio2023,multifuzz2024,hoedur2023}. While these approaches allow for simpler and potentially faster emulation due to the greatly reduced rule-set to analyse, the cost is generally lower fidelity. The lower fidelity generally results in the executed paths being a superset of those reachable on the hardware. This includes some false positive paths that would be unreachable with more accurate peripheral emulation. Notably, the described approaches focus on MMIO input, and do not include DMA. Efforts to extend these approaches to emulate DMA are an ongoing research area, but existing runtime analysis approaches are limited in their ability to identify some types of DMA~\cite{dice2021}.

\section{DMA Controller Study}\label{sec:dma-study}
In its simplest form, a DMA controller copies data from a specified source to a destination. Essentially, the DMA controller performs a task similar to the C standard function \texttt{memcpy()}, reading data from one location and storing it at another. However, microcontrollers integrate DMA controllers with internal peripheral devices. This makes the DMA controller aware of when data is delivered to peripherals, and it can transparently move this data to system memory without intervention from the CPU. The DMA controller can then generate a single event to inform the CPU that a large chunk of data has been transferred. For cases that transfer data from the peripheral to system memory, the CPU can now process this entire chunk; for cases transferring data to peripherals, the CPU can now reconfigure the DMA controller to send the next chunk of data. By removing the requirement for CPU intervention on every byte of data received, clock cycles are freed for the processor to perform other tasks.

Despite the common inclusion of DMA controllers to MCUs, implementations vary greatly by manufacturer or product line, leading to many different internal layouts. Some products contain multiple different DMA interfaces in a single microcontroller. We perform an analysis of many implementations from Cortex-M based devices sourced from the supported platform list of RiotOS~\cite{riotos_cpus}, a popular realtime operating system supporting more than 200 boards. We inspect MCUs from four popular manufacturers where RiotOS supports multiple products at differing performance levels and analyze the provided interfaces. We gather information based on manual exploration of the relevant datasheets, where we locate DMA-enabled peripherals by searching for the terms: ``DMA", ``Pointer" and ``Bus Master". We record any structures used to configure DMA, and the locations of these structures. We discuss the findings and knowledge synthesized in the following sections.

\def\arraystretch{1.2}
\begin{table}[h]
\centering
\caption{Summary of DMA Interface Analysis}
\label{tab:dma_interfaces}
\resizebox{\columnwidth}{!}{
\begin{tabular}{l|c|c|c|c|c|l|c|} 
\cline{2-8}
                  & \textbf{Manufacturer}                                                          & \begin{tabular}[c]{@{}c@{}}\textbf{Product}\\\textbf{Series}\end{tabular} & \begin{tabular}[c]{@{}c@{}}\textbf{Unique}\\\textbf{DMA}\\\textbf{Interfaces}\end{tabular} & \begin{tabular}[c]{@{}c@{}}\textbf{Dedicated }\\\textbf{Controller}\\\textbf{Regions}\end{tabular} & \begin{tabular}[c]{@{}c@{}}\textbf{Unique }\\\textbf{Integrated}\\\textbf{Interfaces}\end{tabular} & \multicolumn{1}{c|}{\begin{tabular}[c]{@{}c@{}}\textbf{Low/High}\\\textbf{end series}\end{tabular}} & \begin{tabular}[c]{@{}c@{}}\textbf{Descriptor}\\\textbf{Locations}\end{tabular}                                                 \\ 
\cline{2-8}
\multirow{2}{*}{} & \multirow{2}{*}{ST}                                            & STM32F0~\cite{stm32f0}                                                                   & 1                                                                                          & 2                                                                                                  & 0                                                                                                  & Low                                                                                                 & MMIO                                                                                                                            \\ 
\hhline{~~------|}
                  &                                                                                & {\cellcolor[rgb]{0.882,0.937,1}}STM32F7~\cite{stm32f7}                                   & {\cellcolor[rgb]{0.882,0.937,1}}5                                                          & {\cellcolor[rgb]{0.882,0.937,1}}2                                                                  & {\cellcolor[rgb]{0.882,0.937,1}}4                                                                  & {\cellcolor[rgb]{0.882,0.937,1}}High                                                                & {\cellcolor[rgb]{0.882,0.937,1}}\begin{tabular}[c]{@{}>{\cellcolor[rgb]{0.882,0.937,1}}c@{}}MMIO, RAM,\\Integrated\end{tabular}  \\ 
\cline{2-8}
\multirow{2}{*}{} & \multirow{2}{*}{\begin{tabular}[c]{@{}c@{}}Nordic\\Semiconductor\end{tabular}} & nRF51~\cite{nrf51}                                                                     & 5                                                                                          & 0                                                                                                  & 5                                                                                                  & Low                                                                                                 & Integrated                                                                                                                      \\ 
\hhline{~~------|}
                  &                                                                                & {\cellcolor[rgb]{0.882,0.937,1}}nRF53~\cite{nrf53}                                     & {\cellcolor[rgb]{0.882,0.937,1}}13                                                         & {\cellcolor[rgb]{0.882,0.937,1}}0                                                                  & {\cellcolor[rgb]{0.882,0.937,1}}13                                                                 & {\cellcolor[rgb]{0.882,0.937,1}}High                                                                & {\cellcolor[rgb]{0.882,0.937,1}}Integrated                                                                                      \\ 
\cline{2-8}
\multirow{2}{*}{} & \multirow{2}{*}{Microchip}                                                     & SAM L10~\cite{saml10}                                                                   & 1                                                                                          & 1                                                                                                  & 0                                                                                                  & Low                                                                                                 & RAM                                                                                                                             \\ 
\hhline{~~------|}
                  &                                                                                & {\cellcolor[rgb]{0.882,0.937,1}}SAM 4S~\cite{sam4s}                                    & {\cellcolor[rgb]{0.882,0.937,1}}1                                                          & {\cellcolor[rgb]{0.882,0.937,1}}0                                                                  & {\cellcolor[rgb]{0.882,0.937,1}}1                                                                  & {\cellcolor[rgb]{0.882,0.937,1}}High                                                                & {\cellcolor[rgb]{0.882,0.937,1}}Integrated                                                                                      \\ 
\cline{2-8}
\multirow{2}{*}{} & \multirow{2}{*}{\begin{tabular}[c]{@{}c@{}}Texas\\Instruments\end{tabular}}    & CC2538~\cite{cc2538}                                                                    & 1                                                                                          & 1                                                                                                  & 0                                                                                                  & Low                                                                                                 & RAM                                                                                                                             \\ 
\hhline{~~------|}
                  &                                                                                & {\cellcolor[rgb]{0.882,0.937,1}}CC26X0~\cite{cc26x0}                                    & {\cellcolor[rgb]{0.882,0.937,1}}2                                                          & {\cellcolor[rgb]{0.882,0.937,1}}1                                                                  & {\cellcolor[rgb]{0.882,0.937,1}}1                                                                  & {\cellcolor[rgb]{0.882,0.937,1}}High                                                                & {\cellcolor[rgb]{0.882,0.937,1}}MMIO, RAM                                                                                       \\
\hhline{~-------|}
\end{tabular}
}
\end{table}
\def\arraystretch{1}

\subsection{Controller Implementations}
We observe that not all DMA interfaces are represented by a dedicated DMA peripheral region. Instead, some DMA interfaces are integrated into other peripheral's MMIO regions. Furthermore, some DMA interfaces are located outside of MMIO entirely. Instead, they elect to configure DMA within a pre-defined structure stored in RAM, and load this DMA configuration into the DMA controller using DMA. Table~\ref{tab:dma_interfaces} shows a summary of information gathered from manufacturer datasheets regarding the number of unique DMA interfaces, and their locations for different microcontrollers.

\subsubsection{Multi-Channel DMA Controller ($I_M$)}\label{sec:multichannel}
A simple implementation of a DMA controller includes a memory mapped region containing two pointers, control registers for configuration, and status registers for the state of the current transfer. This grouping forms a transfer descriptor. Often, both the \textit{source pointer} and \textit{destination pointer} are explicitly defined by the programmer and written to MMIO addresses within the DMA controller. Configuration options such as how many values should be transferred, how large a value is, or the direction for the transfer are stored within the control registers. Figure~\ref{fig:multichannel}~(a) shows a simplified configuration using this structure, based on the STM32F0 series. We observe several unrelated values (\textit{SIZE}, \textit{DIR} and \textit{EN} in the example) packed into adjacent bits within one register. Information regarding the state of DMA transfers, such as whether a transfer is complete or how many values have been transferred, is stored within status registers. Due to the lack of a standard transfer descriptor format, the order of the registers, and any fields contained within, varies.

To handle the large number of peripherals integrated into an MCU, a DMA controller is usually extended to operate with multiple \textit{channels}. Each channel has its own transfer descriptor. This allows for multiple transfers to be active simultaneously, each with their own source and destination pointers.  Different products may have varying numbers of channels available.

\subsubsection{Peripheral Integrated DMA Controller ($I_I$)}\label{sec:integrated}
Not all DMA controllers are accessed through a dedicated interface. Some implementations rely on configuring DMA for each peripheral within the peripheral's own MMIO region. Since the DMA controller is integrated with the peripheral, the descriptor is simplified. This transfer descriptor may only contain a single pointer---the RAM pointer---being read or written. The peripheral pointer for a DMA transfer is \textit{not} set up by the programmer, but is selected implicitly. Other information such as the size of each value being transferred can also be implicitly known based on the peripheral, rather than being defined by the programmer. An example of this integrated interface type for the nRF51 series is shown in Figure~\ref{fig:integrated}~(b).

\subsubsection{List-based DMA Controller ($I_R$)}\label{sec:listbased}

Not all DMA transfer descriptors are stored in MMIO. Some DMA controllers elect to read transfer descriptors using DMA. Microcontrollers such as the Microchip SAM L10 series use a list-based structure in RAM to set up DMA. A single pointer to the head of a list of DMA transfer descriptors is passed to the DMA controller, and it iterates through each entry and sets up the appropriate transfers. An example is shown in Figure~\ref{fig:listbased}~(c).

While the example uses a linked list, we also observe implementations such as the CC2538 using arrays. Instead of using a pointer to the next list entry, the DMA transfer descriptors are stored in contiguous memory. This example has the descriptor list address within a dedicated DMA controller region, notably, some MCUs may use list-based descriptor lists within an integrated interface.

\subsubsection{Varied Interfaces within Product Families}\label{sec:varied_interfaces}
While the previously described examples of DMA controllers are based on a variety of different manufacturers, we do not find the use of one structure consistent within that manufacturer, or even a product family. Although the Microchip SAM L10 uses a list-based structure, the higher performance SAM 4S integrates DMA into peripherals ($I_I$).

The STM32F7 microcontrollers do include a dedicated DMA controller MMIO region, but also contains additional interfaces for peripherals such as USB or Ethernet. USB utilises an integrated DMA interface, comparable to $I_I$, while the Ethernet peripheral uses an integrated DMA controller with descriptors stored in RAM, similar to $I_R$. The dedicated DMA controller region, also differs from the STM32F0, containing additional features, including the addition of a second memory pointer for each transfer descriptor.

Nordic Semiconductor's newer nRF53 series utilises similar interfaces to the nRF51 series, but adds DMA integration to many more peripherals, with different layouts used for DMA transfer descriptors. One notable difference between these interfaces is the definition of the buffer size. While several peripherals have a register dedicated to setting the maximum buffer size, some have pre-defined buffer sizes that are always used for that peripheral, thus it is not explicitly defined by any configuration register.

We observe similar layouts with minor variations from Texas Instruments. Their µDMA controller stores DMA descriptors in RAM, but uses a contiguous array of descriptors, rather than a linked list. While this interface was used in both MCUs from this manufacturer, we also observed the newer series included a second, integrated DMA interface for one peripheral.

DMA controller interfaces are numerous and constantly evolving as more products are released, making a comprehensive list difficult to maintain. Given the inconsistency of DMA interface designs, the key challenge is to devise analysis techniques that can generalise to different target MCUs. We develop  an approach based around the only consistently present component---a pointer to RAM---in the following section.

\vspace{1mm}
\begin{mdframed}
[backgroundcolor=blue!10,
linecolor=blue!60!black,
linewidth=1pt,
roundcorner=2mm] 
    \textbf{Key Takeaway}: Key information used to define a DMA transfer is packed into a structure: a DMA transfer descriptor. But the location and internal layout of descriptors vary significantly across DMA controller implementations. While peripheral pointers and transfer sizes are often ambiguous or implicitly defined, a pointer to a buffer in RAM is the only component consistently observed across DMA interfaces. The unknowns make it difficult to accurately identify the location and size of DMA transfers without external information. 
\end{mdframed}

\section{\name{} Design}\label{design}
\begin{figure}[h!]
    \centering
    \includegraphics[width=\columnwidth]{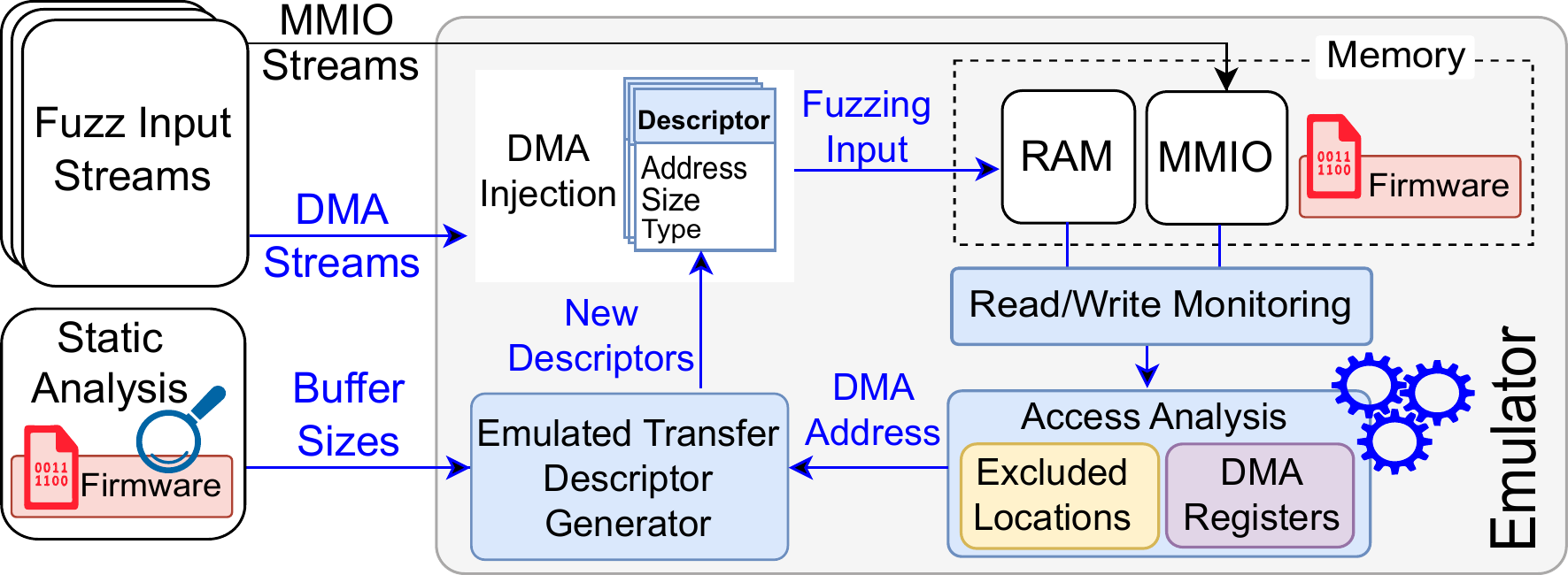}
    \caption{Overview of \name{}. We combine memory analysis techniques with data from static analysis to generate emulated transfer descriptors for fuzzer input streams to inject data into DMA peripherals.}
    \label{fig:overview}
\end{figure}

An overview of our approach, \name{}, is shown in Figure~\ref{fig:overview}. To fuzz inputs to DMA-enabled peripherals, we must identify the DMA buffers that data is delivered to. This information is specified within DMA transfer descriptors used for data input. We monitor RAM and MMIO use in \textit{Access Analysis} to detect components of DMA descriptors---particularly pointers to DMA buffers or in-memory DMA descriptors, illustrated in Figure~\ref{fig:listbased}---within the manufacturer defined layouts and combine this information with buffer sizes obtained from static analysis within the \textit{Emulated Transfer Descriptor Generator} to create emulated transfer descriptors with a consistent format. The emulator can interpret these Descriptors to facilitate DMA data injection. In the following we describe the:
\begin{itemize}[itemsep=2pt,parsep=0pt,topsep=3pt,labelindent=0pt,leftmargin=5mm]
    \item Heuristics we developed to identify MMIO registers related to the DMA transfer descriptors (Section~\ref{identify_registers}) and the resultant buffer (Section~\ref{size_inference}).
    \item The inference techniques to detect DMA types and extract in-memory DMA descriptors (Section~\ref{identify_type}). 
    \item Then, the generation of emulated DMA transfer descriptors, to realize a consistent interface (Section~\ref{emulated_transfer}) the emulator can interpret, allowing for data injection (Section~\ref{inject_data}). 
\end{itemize}

\subsection{Identifying DMA Transfers}

A key component of identifying firmware DMA buffers is identifying the locations that are used to store DMA transfer descriptors. Existing approaches either gather this information based on manufacturer supplied information~\cite{semu2022,perry2024}, or use values written to adjacent MMIO locations as a heuristic to identify pointers at runtime~\cite{dice2021}. In contrast, our approach focuses on identifying the memory addresses used within DMA transfer descriptors, the size of the buffer, and the type of DMA descriptors used, without assumptions of the DMA descriptor's layout.

\begin{figure}[h!]
    \centering
    \includegraphics[width=\columnwidth,trim={0 0.48cm 0 0.04cm},clip]{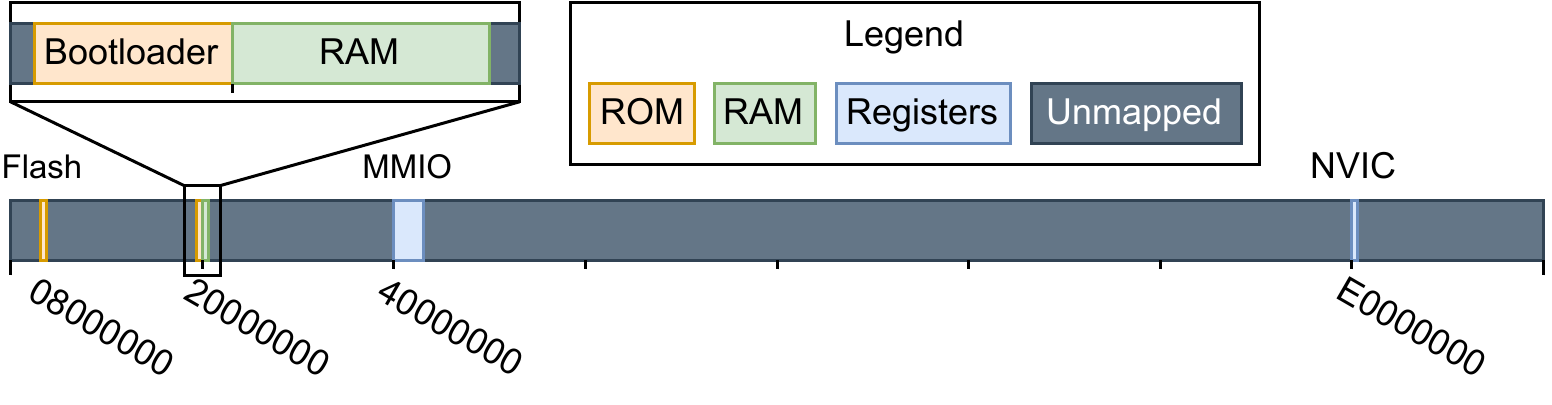}
    \caption{An example memory map demonstrating the limited range of RAM values within the address-space.}
    \label{fig:memorymap}
\end{figure}

\subsubsection{DMA Address Registers}\label{identify_registers}
As described in Section~\ref{sec:dma-study}, a destination memory address is a consistently present part of a DMA descriptor, while we observe other components to either be implicitly defined \Circled{C2}, or packed into a single memory location with other variables. We start with the assumption that any MMIO address could contain pointers used to configure DMA, and incrementally refine knowledge of DMA related registers across the many different executions from different fuzzer generated test cases. By analysing the values written to each location, and the corresponding access sizes, we classify MMIO registers based on whether they consistently contain pointers within a DMA transfer descriptor.

\noindent\textbf{Memory Values.~}Our approach is motivated by the knowledge that valid pointers to RAM are rarely written to MMIO. Figure~\ref{fig:memorymap} shows a simplified version of the STM32F0~\cite{stm32f0} series' memory map. Modern 32-bit MCUs typically only contain kilobytes of RAM. Consequently, only a tiny fraction (less than 0.001\% for the STM32F0) of the possible values writable to MMIO are valid RAM pointers.

Any MMIO registers that have non-pointer values written to them are excluded from consideration for DMA address registers. The vast majority of MMIO registers, with pseudo-random numbers written to them, can be excluded from consideration for DMA based on the first write alone. However, fuzzers generate millions of test cases, and fuzzer inputs can often control the values written to MMIO. We expect that fuzzers would eventually discover inputs that write pointer-like values to MMIO. To resolve this source of potential false-positives, we extend the state information for DMA analysis beyond a single executed test case, tracking the written values across \textbf{\textit{all}} executed test cases. Once a non-pointer value is written to an MMIO register, that MMIO register is not considered for DMA in any future test cases. Contrastingly, when registers that have not been excluded receive a valid pointer value, they are considered a basis for an \textit{emulated DMA transfer}.

\vspace{1mm}
\noindent\textbf{Access Size.~}A pointer is a word-sized object. Accessing a location with a non-word-sized read or write indicates the value stored is not a pointer, even if a sequence of writes could create a pointer-like value. We augment our memory value based analysis to exclude non-word-sized memory locations from consideration for DMA pointers.

\subsubsection{Identifying DMA Buffers.~}\label{size_inference}During execution, we expect that eventually a valid RAM pointer will be written to a DMA pointer configuration register previously identified using the method in Section~\ref{identify_registers}. Once this occurs, we need to determine the buffer size before we can accurately inject fuzzing data.

\vspace{1mm}
\noindent\textbf{DMA Buffer Size Inference.~}
The lack of a standardized DMA descriptor \Circled{C1} makes determining how many bytes of data the DMA controller needs to transfer challenging. Extracting the size from the descriptor would require identifying which word of memory within the descriptor contains the transfer count, which is not possible when sizes are implicitly defined \Circled{C2}. The units used to define the size aren't consistent; it could be bytes, words or half-words. The problem is compounded by the fact that the size can be packed with other variables into one value, as shown in Figure~\ref{fig:multichannel}~(a). Without MCU-specific information to define these variables, we must find an alternate method to infer buffer sizes.

Rather than extracting the size of the DMA transfer directly, we identify the size of the target buffer. We leverage the open-source static analysis tool Ghidra~\cite{ghidra} to assist in identifying the size of DMA buffers based on static analysis. We extract variables, arrays, and structs stored in global variables where available, and use Ghidra's reference detection to mark boundaries when in-depth information is missing. This information is imported into \name{} at start-up.
If a buffer is discovered at run-time, where no information was gathered using Ghidra, we fall back to a method that increments the buffer size as sequential memory addresses are accessed with a consistent size, based on that used by DICE~\cite{dice2021}.

\subsubsection{Identifying DMA Type}\label{identify_type}
The varied DMA descriptor locations pose a further problem \Circled{C3}. As described in Section~\ref{sec:listbased}, descriptors may not always point to the target buffers. Instead some DMA controller implementations are configured from RAM, and utilize DMA to read \textit{in-memory DMA transfer descriptors}. Lists of DMA transfer descriptors in RAM ($I_R$) need to be modeled separately from the more typical DMA buffers pointed to directly from a DMA transfer descriptor in MMIO ($I_M$ and $I_I$). Rather than fuzzing the list contents, the list is analysed and each transfer descriptor configured within is extracted and emulated.

We can identify whether a pointer points to a DMA buffer, or a DMA transfer descriptor list entry based on the content stored inside the object. While we have no prior information available to identify configuration or status values, we know the memory map, allowing us to identify pointer values. DMA descriptors notably contain valid pointers. Figure~\ref{fig:listbased}~(c) shows each transfer's descriptor contains at least one pointer to another variable in RAM. However, we would not expect to observe pointer values within a buffer intended for sending or receiving peripheral data, for example, the \textit{SPI Input Buffer} in Figure~\ref{fig:multichannel}~(a) would be zeroed. We consider this buffer similar to another MMIO region, and apply the rules described in Section~\ref{identify_registers} to search for DMA addresses. We note that in-memory DMA transfer descriptors may be stored as an array, or a linked list. 
As shown in the example from Figure~\ref{fig:listbased}~(c), linked lists of in-memory DMA transfer descriptors contain a value outside of the typical status, configuration and destination variables; a pointer to an additional DMA transfer descriptor, \textit{List Next}. We must determine whether each pointer points to a DMA buffer, or another list entry in the DMA transfer descriptor list.
Once our method identifies a DMA list entry, the logic is recursively used to classify each buffer pointed to from within the parent buffer. To handle cases where the configuration is still being written, we continue to monitor writes to potential list entries, until a read occurs. Once the object is read, we set the type to a \textit{DMA transfer descriptor} if a \textbf{pointer was identified}, or a \textit{DMA buffer} if \textbf{no pointer was found} within the buffer. DMA list information is propagated back to the originating MMIO address for future reference.

\subsection{Emulated DMA Transfer Descriptor}\label{emulated_transfer}
In order to provide a consistent interface to interact with the emulator, we created an emulated DMA transfer descriptor. This structure encapsulates information regarding the \textbf{address}, \textbf{size}, \textbf{type} and \textbf{accesses} for a DMA transfer. The address represents the starting address of the target DMA buffer. Size represents the size of the target DMA buffer, or 0 if the size is unknown. The type indicates, if known, whether a buffer is designated as an in-memory DMA transfer descriptor entry as part of a list, or a target DMA buffer. Information on prior accesses is recorded, and serves a dual purpose: i)~the recorded information is required to perform type inference (Section~\ref{identify_type}) on an unclassified object; and ii)~it serves as a source of history for the fall back method (as described in Section~\ref{size_inference}) when the buffer size cannot be statically identified. The history of accesses maintains knowledge of the required size and the next valid address in the sequence. Notably, each emulated transfer descriptor instance can have a distinct type, allowing for multiple DMA descriptor layouts on a single MCU~\Circled{C5}, and is generic, encapsulating DMA transfer information for the ever growing number of unique DMA interfaces~\Circled{C4}.

\subsection{Injecting DMA Data}\label{inject_data}
After identifying the DMA transfer descriptor destinations, we need to be able to insert data into the buffer. We perform data injection into DMA buffers just-in-time, mutating the value as the firmware attempts to read data inside the buffer. When the buffer content is read, data from an external source (in this case a fuzzer) is inserted in place of the existing data stored at that location.

We note that DMA transfer descriptors stored in memory require special consideration. The DMA controller has the capability to modify the structures in memory to indicate the state of the transfer. The DMA controller is responsible for informing the firmware of state information such as whether the transfer has been completed. For descriptors in MMIO, the associated status values will be fuzzed as they are within the existing scope of the fuzzer.  To allow fuzzing these values for descriptors located in RAM, we expand the scope of the fuzzer to allow fuzzing some values within the descriptor. Notably, we exclude the pointers identified within the descriptor from being fuzzed, as mutating pointers would corrupt the system's state. Additionally, the detected DMA transfer descriptors include both DMA based input and output. However, DMA based outputs are not fuzzed, as the just-in-time read method is never triggered for output buffers, as they never read by the CPU.

\section{Implementation}
We built a prototype implementation of our approach on top of MultiFuzz~\cite{multifuzz2024}, a state-of-the-art multi-stream fuzzer for monolithic embedded firmware. This approach allows stream specific mutation strategies and the application of fuzzing techniques to solve roadblocks such as \textit{magic bytes}~\cite{redqueen2019}. As well as benefiting from the performance advantages of multi-stream fuzzing techniques, the underlying emulator \icicle~\cite{icicle2023}, is modern, extensible, and built with a focus on instrumentation and execution of firmware.

Memory hook interfaces for \icicle~\cite{icicle2023} are used to track reads and writes to both RAM and MMIO regions. We implement the logic described in Section~\ref{design} within these memory hooks. When a DMA buffer is read, we read from a fuzzing stream identified by the buffer's base address. To minimize performance overheads, we initially start with no hooks on RAM addresses, only generating the first hook once a DMA buffer is identified. This region dynamically grows to cover all memory from the start of the first DMA buffer to the end of the last DMA buffer in memory. The performance impact of our hooks is discussed further in Section~\ref{sec:overhead}.

The pointers stored within DMA transfer descriptors are persistent (i.e. should not be fuzzed when they are read). This goes against the default behaviour for many monolithic fuzzers. To address this, we prevent MMIO registers detected as DMA transfer pointers from being fuzzed, and treat them as memory. Consequently, changing a register model can modify the behaviour of a fuzz input; and may retroactively influence seeds discovered earlier in the fuzzing campaign. To maintain consistency after models are updated, we implement additional logic applied after each model change: we clear all coverage information, re-queue all previously discovered seeds, and clear the stored seeds. As the queued seeds are executed, coverage information is recomputed, and interesting seeds will be preserved, while seeds that no longer reach interesting code paths are discarded.

\section{Evaluation}
To evaluate the effectiveness of \name{}, we conduct a methodical evaluation of our fuzzer, in stages, and compare with state-of-the-art approaches for injecting fuzzing data into emulated firmware that employ DMA for obtaining data from peripherals. These experiments aim to answer the following research questions:

\begin{itemize}[itemsep=1pt,parsep=0pt,topsep=3pt,labelindent=4pt,leftmargin=8mm]
 \item[\textbf{RQ1}] How does \name{} compare with the state-of-the-art DMA fuzzers in detecting DMA configurations across a wide range of microcontrollers and peripherals? (Section~\ref{sec:unittests})
 \item[\textbf{RQ2}] How effective is \name{} in exploring code compared to state-of-the-art fuzzers? (Section~\ref{coverage})

 \item[\textbf{RQ3}] How does \name{} compare in bug-finding capability and influence the ability to discover bugs in real-world firmware samples? 
 (Section~\ref{bug_replication})
 \item[\textbf{RQ4}] How does our dynamic analysis techniques impact test throughput during execution?  (Section~\ref{sec:overhead})
\end{itemize}

\noindent\textbf{Fuzzers.~}To evaluate the impact of \name, we compare with current state-of-the-art fuzzers. We select MultiFuzz~\cite{multifuzz2024} (Baseline with no support DMA) since \name{} builds upon its stream-based input representation. Further, we re-implemented DICE's~\cite{dice2021} DMA detection onto MultiFuzz as DICE was originally implemented atop P2IM without support for multi-stream inputs, lowering fuzzing performance. This re-implementation allows us to isolate the DMA impact, separate from the influence of the underlying frameworks. We also consider DICE~\cite{dice2021} and SEmu~\cite{semu2022} where possible.

\vspace{1mm}
\noindent\textbf{Target Binaries.~}We evaluate the DMA capabilities of our fuzzer with an extensive set of \targets{} binaries including a mixture of multi-channel ($I_M$), integrated ($I_I$) and list-based $(I_R)$ DMA interfaces. This set contains binaries built for 11 product families across 6 manufacturers. 33\% of these manufacturers and 82\% of these product families are outside of the set considered in Section~\ref{sec:dma-study} analysis; details of individual binaries and corresponding product families are in our git repository~\cite{gitrepobinaries}.
\begin{enumerate}
[itemsep=2pt,parsep=0pt,topsep=3pt,labelindent=0pt,leftmargin=5mm]
    \item All arm-based DMA \& non-DMA unit tests from DICE~\cite{dice2021} \& P2IM~\cite{p2im2020}: \textbf{80}
    \item All DMA enabled open-source firmware binaries from DICE~\cite{dice2021} and SEmu~\cite{semu2022}'s evaluation: \textbf{11}
    \item New or updated binaries compiled to include DMA from open-source projects in prior non-DMA studies: \textbf{3}
    \item Bug-based evaluation with \textbf{8} CVEs where the bugs are triggered by data injected from DMA peripherals. Sourced from Fuzzware~\cite{fuzzware2022} where we removed the manual code patches applied to enable the binaries to execute in previous studies.
    \item Non-DMA binaries from P2IM~\cite{p2im2020} to evaluate overheads: \textbf{6}
\end{enumerate}

\noindent\textbf{Configurations and Settings.~}To allow sufficient time to trigger existing bugs known to be difficult to discover, we use 96 hour trials, with 10 trials per binary for benchmarking, unless stated otherwise. Additionally, longer trials allow analysis of potential overheads incurred by \name{'s} runtime monitoring.
We follow established fuzzing guidelines~\cite{fuzzingsok2024,fuzzingsokrepo2024} for our benchmarking evaluations. A single CPU core from an AMD Threadripper Pro 5995WX is allocated to each fuzzing instance in a virtualized environment. Each fuzzer is given an identical or equivalent configuration for each tested binary where applicable. In the case of SEmu~\cite{semu2022}, we use existing MCU models provided by the authors for supported targets; unsupported targets are not run. Code coverage results for the original DICE~\cite{dice2021} implementation are omitted as our reimplementation on MultiFuzz showed higher coverage in every case.

\subsection{Dynamic DMA Model Analysis Veracity}\label{sec:unittests}
To indicate the capabilities of the approach in comparison to the existing state-of-the-art runtime DMA modeling methods, we investigate the reported DMA buffers in the unit test binaries set from DICE~\cite{dice2021} and P2IM~\cite{p2im2020}. The set consists of 32 DMA binaries targeting 11 different product lines across 5 different manufacturers, and 48 non-DMA binaries targeting 5 different MCUs. For each binary where a DMA buffer was identified, we manually check the MMIO register associated with this DMA transfer in the manufacturer's datasheet. In cases where the manufacturer indicates the register is a pointer for a DMA enabled peripheral, we consider this a true positive. If any detected buffers are associated with an MMIO register that is not designated for use as a DMA pointer, we classify it as a false positive. The \textit{NRF51822\_SPI\_slave} from the DICE unit tests is moved to the \textit{Binaries without DMA} set, as manual analysis shows it uses MMIO access for its transfers to the SPI peripheral, not DMA. As \name{} generates its models based on several generated test cases with varied input values from a fuzzer, we allow \name{} a single fuzzing run with a 5 minute limit to identify the models for each unit test.

\begin{table}[h!]
\centering
\caption{DMA Identification Accuracy. Number of binaries detected to contain DMA, grouped by true/false positives.}
\label{tab:unit_tests}
\resizebox{\columnwidth}{!}{%
\begin{tabular}{|l|l|l|l|} 
\cline{3-4}
\multicolumn{1}{l}{}                                      &                & \name      & \dice{}~\cite{dice2021}    \\ 
\hline
\multirow{2}{*}{\textbf{\textbf{Binaries with DMA (32)}}} & True Positive  & 32 (100\%) & 26 (81\%)  \\ 
\cline{2-4}
                                                          & False Positive & 0 (0\%)    & 0 (0\%)    \\ 
\hline
\textbf{\textbf{Binaries without DMA (48)}}               & False Positive & 0 (0\%)    & 0 (0\%)    \\
\hline
\end{tabular}
}
\end{table}

As shown in Table~\ref{tab:unit_tests}, \name{} can effectively identify DMA buffers across several device manufacturers and product lines without introducing any false positives across the test suite. Compared to an existing state-of-the-art approach, DICE~\cite{dice2021}, \name{} was able to identify 6 DMA configurations missed by DICE. Manual analysis shows these 6 cases were the result of DICE's heuristics missing an integrated DMA interface ($I_I$), where no peripheral pointer register is present.

\definecolor{Jumbo}{rgb}{0.494,0.494,0.501}
\def\arraystretch{1.45}
\begin{table*}[t!]
\caption{Block coverage achieved for each binary and fuzzer after 96 hours. Relevant DMA implementation types for each binary are shown in brackets. Cases Marked N/A designates that the fuzzer does not currently support the target microcontroller. Results marked with an asterisk contain memory corruption influencing coverage results. P values are presented comparing each fuzzer to \name{}, based on Mann-Whitney U tests, at a 0.01 significance level.}
\label{fig:cov_tab}
\centering
\resizebox{\textwidth}{!}{%
\begin{tabular}{|c|r|r|c|c|c|c|c|c|c|c|c|c|c|c|c|c|c|} 
\cline{4-18}
\multicolumn{3}{l|}{}                                                                                                                                 & \multicolumn{4}{c|}{\textbf{SEmu}}                                                                                                                                                                                       & \multicolumn{4}{c|}{\textbf{MultiFuzz}}                                                                                                                                                                         & \multicolumn{4}{c|}{\textbf{MultiFuzz-DICE}}                                                                                                                                                                                      & \multicolumn{3}{c|}{\textbf{\name}}                                                                                                                                    \\ 
\hline
\multicolumn{2}{|r|}{\textbf{DMA Binary Dataset}}                                                            & \textbf{\#Blocks}                      & \multicolumn{1}{r|}{\textbf{Min}}                     & \textbf{Med}                                          & \textbf{Max}                                 & \textit{p}                                                & \textbf{Min}                                          & \textbf{Med}                                 & \textbf{Max}                                 & \textit{p}                                                & \textbf{Min}                                          & \textbf{Med}                                          & \textbf{Max}                                          & \textit{p}                                                & \textbf{Min}                                          & \textbf{Med}                                          & \textbf{Max}                                           \\ 
\hline
\multirow{7}{*}{\begin{sideways}DICE~\cite{dice2021}\end{sideways}} & GPS Receiver ($I_M$)                                       & 3006                                   & \hl{11.8\%}                                           & \hl{11.8\%}                                           & \hl{11.8\%}                                  & \textbf{\textless{}0.01}                                  & \hl{12.1\%}                                           & \hl{12.1\%}                                  & \hl{12.1\%}                                  & \textbf{\textless{}0.01}                                  & \hl{35.2\%}*                                          & \hl{43.4\%}*                                          & \hl{47.6\%}*                                          & N/A                                                       & \textbf{\hl{43.4\%}}                                  & \textbf{\hl{44.1\%}}                                  & \textbf{\hl{45.6\%}}                                   \\
                                                & {\cellcolor[rgb]{0.89,0.89,0.89}}Guitar Pedal ($I_M$)      & {\cellcolor[rgb]{0.89,0.89,0.89}}8213  & {\cellcolor[rgb]{0.89,0.89,0.89}}N/A                  & {\cellcolor[rgb]{0.89,0.89,0.89}}N/A                  & {\cellcolor[rgb]{0.89,0.89,0.89}}N/A         & {\cellcolor[rgb]{0.89,0.89,0.89}}N/A                      & {\cellcolor[rgb]{0.89,0.89,0.89}}\textbf{\hl{27.2\%}} & {\cellcolor[rgb]{0.89,0.89,0.89}}\hl{27.2\%} & {\cellcolor[rgb]{0.89,0.89,0.89}}\hl{27.3\%} & {\cellcolor[rgb]{0.89,0.89,0.89}}0.12                     & {\cellcolor[rgb]{0.89,0.89,0.89}}\hl{27.1\%}          & {\cellcolor[rgb]{0.89,0.89,0.89}}\hl{27.2\%}          & {\cellcolor[rgb]{0.89,0.89,0.89}}\hl{27.3\%}          & {\cellcolor[rgb]{0.89,0.89,0.89}}\textbf{\textless{}0.01} & {\cellcolor[rgb]{0.89,0.89,0.89}}\textbf{\hl{27.2\%}} & {\cellcolor[rgb]{0.89,0.89,0.89}}\textbf{\hl{27.3\%}} & {\cellcolor[rgb]{0.89,0.89,0.89}}\textbf{\hl{27.4\%}}  \\
                                                & MIDI Synth ($I_M$)                                         & 703                                    & \hl{20.6\%}                                           & \hl{20.6\%}                                           & \hl{20.6\%}                                  & \textbf{\textless{}0.01}                                  & \hl{29.6\%}                                           & \hl{29.6\%}                                  & \hl{29.6\%}                                  & \textbf{\textless{}0.01}                                  & \textbf{\hl{46.8\%}}                                  & \textbf{\hl{46.8\%}}                                  & \textbf{\hl{46.8\%}}                                  & N/A                                                       & \textbf{\hl{46.8\%}}                                  & \textbf{\hl{46.8\%}}                                  & \textbf{\hl{46.8\%}}                                   \\
                                                & {\cellcolor[rgb]{0.89,0.89,0.89}}Modbus ($I_M$)            & {\cellcolor[rgb]{0.89,0.89,0.89}}810   & {\cellcolor[rgb]{0.89,0.89,0.89}}N/A                  & {\cellcolor[rgb]{0.89,0.89,0.89}}N/A                  & {\cellcolor[rgb]{0.89,0.89,0.89}}N/A         & {\cellcolor[rgb]{0.89,0.89,0.89}}N/A                      & {\cellcolor[rgb]{0.89,0.89,0.89}}\hl{49.4\%}          & {\cellcolor[rgb]{0.89,0.89,0.89}}\hl{49.4\%} & {\cellcolor[rgb]{0.89,0.89,0.89}}\hl{49.4\%} & {\cellcolor[rgb]{0.89,0.89,0.89}}\textbf{\textless{}0.01} & {\cellcolor[rgb]{0.89,0.89,0.89}}\textbf{\hl{58.9\%}} & {\cellcolor[rgb]{0.89,0.89,0.89}}\textbf{\hl{58.9\%}} & {\cellcolor[rgb]{0.89,0.89,0.89}}\textbf{\hl{58.9\%}} & {\cellcolor[rgb]{0.89,0.89,0.89}}N/A                      & {\cellcolor[rgb]{0.89,0.89,0.89}}\textbf{\hl{58.9\%}} & {\cellcolor[rgb]{0.89,0.89,0.89}}\textbf{\hl{58.9\%}} & {\cellcolor[rgb]{0.89,0.89,0.89}}\textbf{\hl{58.9\%}}  \\
                                                & Oscilloscope ($I_M$)                                       & 3199                                   & \hl{5.6\%}                                            & \hl{5.6\%}                                            & \hl{5.6\%}                                   & \textbf{\textless{}0.01}                                  & \hl{26.4\%}                                           & \hl{26.4\%}                                  & \hl{26.4\%}                                  & \textbf{\textless{}0.01}                                  & \hl{27.4\%}                                           & \hl{27.4\%}                                           & \hl{27.4\%}                                           & \textbf{\textless{}0.01}                                  & \textbf{\hl{27.5\%}}                                  & \textbf{\hl{27.5\%}}                                  & \textbf{\hl{27.5\%}}                                   \\
                                                & {\cellcolor[rgb]{0.89,0.89,0.89}}Soldering Station ($I_M$) & {\cellcolor[rgb]{0.89,0.89,0.89}}3311  & {\cellcolor[rgb]{0.89,0.89,0.89}}\hl{31.2\%}          & {\cellcolor[rgb]{0.89,0.89,0.89}}\hl{31.2\%}          & {\cellcolor[rgb]{0.89,0.89,0.89}}\hl{31.3\%} & {\cellcolor[rgb]{0.89,0.89,0.89}}\textbf{\textless{}0.01} & {\cellcolor[rgb]{0.89,0.89,0.89}}\hl{47.1\%}          & {\cellcolor[rgb]{0.89,0.89,0.89}}\hl{47.1\%} & {\cellcolor[rgb]{0.89,0.89,0.89}}\hl{47.1\%} & {\cellcolor[rgb]{0.89,0.89,0.89}}\textbf{\textless{}0.01} & {\cellcolor[rgb]{0.89,0.89,0.89}}\hl{47.1\%}          & {\cellcolor[rgb]{0.89,0.89,0.89}}\hl{47.1\%}          & {\cellcolor[rgb]{0.89,0.89,0.89}}\hl{47.1\%}          & {\cellcolor[rgb]{0.89,0.89,0.89}}\textbf{\textless{}0.01} & {\cellcolor[rgb]{0.89,0.89,0.89}}\textbf{\hl{50.9\%}} & {\cellcolor[rgb]{0.89,0.89,0.89}}\textbf{\hl{51.1\%}} & {\cellcolor[rgb]{0.89,0.89,0.89}}\textbf{\hl{51.1\%}}  \\
                                                & Stepper Motor ($I_M$)                                      & 4009                                   & N/A                                                   & N/A                                                   & N/A                                          & N/A                                                       & \hl{26.2\%}                                           & \hl{26.3\%}                                  & \hl{26.4\%}                                  & \textbf{\textless{}0.01}                                  & \textbf{\hl{37.2\%}}                                  & \textbf{\hl{37.4\%}}                                  & \textbf{\hl{37.8\%}}                                  & 0.17                                                      & \hl{34.6\%}                                           & \hl{37.3\%}                                           & \hl{37.6\%}                                            \\ 
\hline
\multirow{4}{*}{\begin{sideways}SEmu~\cite{semu2022}\end{sideways}} & {\cellcolor[rgb]{0.89,0.89,0.89}}LwIP TCP Client ($I_R$)   & {\cellcolor[rgb]{0.89,0.89,0.89}}5333  & {\cellcolor[rgb]{0.89,0.89,0.89}}\textbf{\hl{29.1\%}} & {\cellcolor[rgb]{0.89,0.89,0.89}}\textbf{\hl{30.6\%}} & {\cellcolor[rgb]{0.89,0.89,0.89}}\hl{33.7\%} & {\cellcolor[rgb]{0.89,0.89,0.89}}0.01                     & {\cellcolor[rgb]{0.89,0.89,0.89}}\hl{26.6\%}          & {\cellcolor[rgb]{0.89,0.89,0.89}}\hl{26.7\%} & {\cellcolor[rgb]{0.89,0.89,0.89}}\hl{26.7\%} & {\cellcolor[rgb]{0.89,0.89,0.89}}\textbf{\textless{}0.01} & {\cellcolor[rgb]{0.89,0.89,0.89}}\hl{26.6\%}          & {\cellcolor[rgb]{0.89,0.89,0.89}}\hl{26.6\%}          & \hl{26.6\%}                                           & {\cellcolor[rgb]{0.89,0.89,0.89}}\textbf{\textless{}0.01} & {\cellcolor[rgb]{0.89,0.89,0.89}}\hl{28.7\%}          & {\cellcolor[rgb]{0.89,0.89,0.89}}\hl{29.2\%}          & {\cellcolor[rgb]{0.89,0.89,0.89}}\textbf{\hl{34.4\%}}  \\
                                                & LwIP TCP Server ($I_R$)                                    & 5212                                   & \hl{26.5\%}                                           & \hl{26.5\%}                                           & \hl{26.7\%}                                  & \textbf{\textless{}0.01}                                  & \hl{26.7\%}                                           & \hl{29.1\%}                                  & \hl{29.1\%}                                  & 0.08                                                      & \hl{26.6\%}                                           & \hl{26.7\%}                                           & \hl{26.7\%}                                           & \textbf{\textless{}0.01}                                  & \textbf{\hl{28.1\%}}                                  & \textbf{\hl{29.3\%}}                                  & \textbf{\hl{35.1\%}}                                   \\
                                                & {\cellcolor[rgb]{0.89,0.89,0.89}}LwIP UDP Client ($I_R$)   & {\cellcolor[rgb]{0.89,0.89,0.89}}4282  & {\cellcolor[rgb]{0.89,0.89,0.89}}\hl{25.2\%}          & {\cellcolor[rgb]{0.89,0.89,0.89}}\hl{26.3\%}          & {\cellcolor[rgb]{0.89,0.89,0.89}}\hl{29.4\%} & {\cellcolor[rgb]{0.89,0.89,0.89}}\textbf{\textless{}0.01} & {\cellcolor[rgb]{0.89,0.89,0.89}}\hl{23.3\%}          & {\cellcolor[rgb]{0.89,0.89,0.89}}\hl{23.4\%} & {\cellcolor[rgb]{0.89,0.89,0.89}}\hl{23.4\%} & {\cellcolor[rgb]{0.89,0.89,0.89}}\textbf{\textless{}0.01} & {\cellcolor[rgb]{0.89,0.89,0.89}}\hl{27.2\%}          & {\cellcolor[rgb]{0.89,0.89,0.89}}\hl{27.2\%}          & {\cellcolor[rgb]{0.89,0.89,0.89}}\hl{27.3\%}          & {\cellcolor[rgb]{0.89,0.89,0.89}}\textbf{\textless{}0.01} & {\cellcolor[rgb]{0.89,0.89,0.89}}\textbf{\hl{30.2\%}} & {\cellcolor[rgb]{0.89,0.89,0.89}}\textbf{\hl{34.9\%}} & {\cellcolor[rgb]{0.89,0.89,0.89}}\textbf{\hl{39.6\%}}  \\
                                                & LwIP UDP Server ($I_R$)                                    & 4135                                   & \hl{21.1\%}                                           & \hl{22.2\%}                                           & \hl{22.5\%}                                  & \textbf{\textless{}0.01}                                  & \hl{20.1\%}                                           & \hl{20.2\%}                                  & \hl{20.2\%}                                  & \textbf{\textless{}0.01}                                  & \hl{20.2\%}                                           & \hl{20.2\%}                                           & \hl{20.2\%}                                           & \textbf{\textless{}0.01}                                  & \textbf{\hl{35.0\%}}                                  & \textbf{\hl{39.3\%}}                                  & \textbf{\hl{40.8\%}}                                   \\ 
\hline
\multirow{3}{*}{\begin{sideways}\name{}~\xspace\end{sideways}}        & {\cellcolor[rgb]{0.89,0.89,0.89}}Contiki HW ($I_R$)        & {\cellcolor[rgb]{0.89,0.89,0.89}}3890  & {\cellcolor[rgb]{0.89,0.89,0.89}}N/A                  & {\cellcolor[rgb]{0.89,0.89,0.89}}N/A                  & {\cellcolor[rgb]{0.89,0.89,0.89}}N/A         & {\cellcolor[rgb]{0.89,0.89,0.89}}N/A                      & {\cellcolor[rgb]{0.89,0.89,0.89}}\hl{25.0\%}          & {\cellcolor[rgb]{0.89,0.89,0.89}}\hl{25.0\%} & {\cellcolor[rgb]{0.89,0.89,0.89}}\hl{25.0\%} & {\cellcolor[rgb]{0.89,0.89,0.89}}\textbf{\textless{}0.01} & {\cellcolor[rgb]{0.89,0.89,0.89}}\hl{25.0\%}          & {\cellcolor[rgb]{0.89,0.89,0.89}}\hl{25.0\%}          & {\cellcolor[rgb]{0.89,0.89,0.89}}\hl{25.0\%}          & {\cellcolor[rgb]{0.89,0.89,0.89}}\textbf{\textless{}0.01} & {\cellcolor[rgb]{0.89,0.89,0.89}}\textbf{\hl{43.1\%}} & {\cellcolor[rgb]{0.89,0.89,0.89}}\textbf{\hl{46.1\%}} & {\cellcolor[rgb]{0.89,0.89,0.89}}\textbf{\hl{47.3\%}}  \\
                                                & Contiki SNMP ($I_R$)                                       & 4549                                   & N/A                                                   & N/A                                                   & N/A                                          & N/A                                                       & \hl{29.0\%}                                           & \hl{29.0\%}                                  & \hl{29.0\%}                                  & \textbf{\textless{}0.01}                                  & \hl{29.0\%}                                           & \hl{29.0\%}                                           & \hl{29.0\%}                                           & \textbf{\textless{}0.01}                                  & \textbf{\hl{50.8\%}}                                  & \textbf{\hl{52.4\%}}                                  & \textbf{\hl{54.4\%}}                                   \\
                                                & {\cellcolor[rgb]{0.89,0.89,0.89}}Pinetime ($I_I$)          & {\cellcolor[rgb]{0.89,0.89,0.89}}31892 & {\cellcolor[rgb]{0.89,0.89,0.89}}N/A                  & {\cellcolor[rgb]{0.89,0.89,0.89}}N/A                  & {\cellcolor[rgb]{0.89,0.89,0.89}}N/A         & {\cellcolor[rgb]{0.89,0.89,0.89}}N/A                      & {\cellcolor[rgb]{0.89,0.89,0.89}}\hl{7.1\%}           & {\cellcolor[rgb]{0.89,0.89,0.89}}\hl{7.1\%}  & {\cellcolor[rgb]{0.89,0.89,0.89}}\hl{7.1\%}  & {\cellcolor[rgb]{0.89,0.89,0.89}}\textbf{\textless{}0.01} & {\cellcolor[rgb]{0.89,0.89,0.89}}\hl{7.1\%}           & {\cellcolor[rgb]{0.89,0.89,0.89}}\hl{7.1\%}           & {\cellcolor[rgb]{0.89,0.89,0.89}}\hl{7.1\%}           & {\cellcolor[rgb]{0.89,0.89,0.89}}\textbf{\textless{}0.01} & {\cellcolor[rgb]{0.89,0.89,0.89}}\textbf{\hl{15.7\%}} & {\cellcolor[rgb]{0.89,0.89,0.89}}\textbf{\hl{15.9\%}} & {\cellcolor[rgb]{0.89,0.89,0.89}}\textbf{\hl{16.0\%}}  \\
\hline
\end{tabular}
}
\vspace{1pt}
\footnotesize
\makecell[l]{
\textit{\textbf{Note}}: The unreachable code included within the binary and its libraries will prevent 100\% coverage from being achievable.}
\end{table*}

\begin{figure*}[h!]
    \includegraphics[width=\textwidth,trim={0 0.4cm 0 0.3cm},clip]{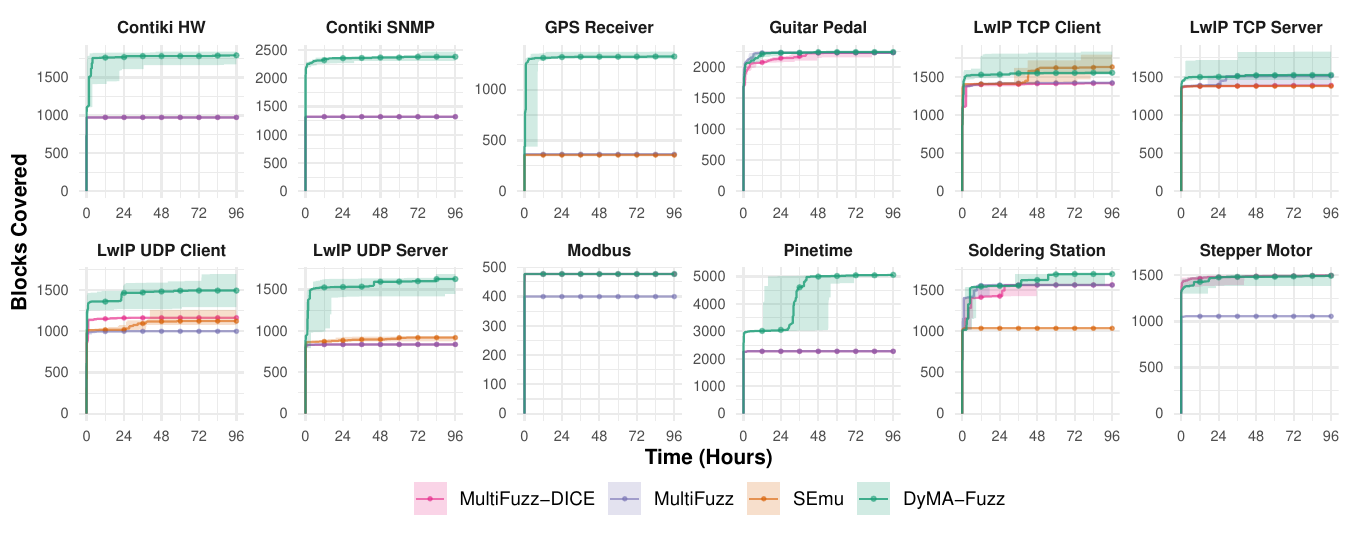}
    \caption{Coverage over time achieved by different approaches. The shaded region indicates the range of coverage results observed across all ten trials. The MIDI Synth and Oscilloscope binaries are omitted for brevity, as more than 99\% of the final coverage for each trial was reached within the first 15 minutes.
    }
    \label{fig:coverage}
\end{figure*}

\subsection{Code Coverage}\label{coverage}

\begin{figure*}[t!]
    \centering
    \includegraphics[width=0.95\linewidth,trim={0 0.3cm 0 0.25cm},clip]{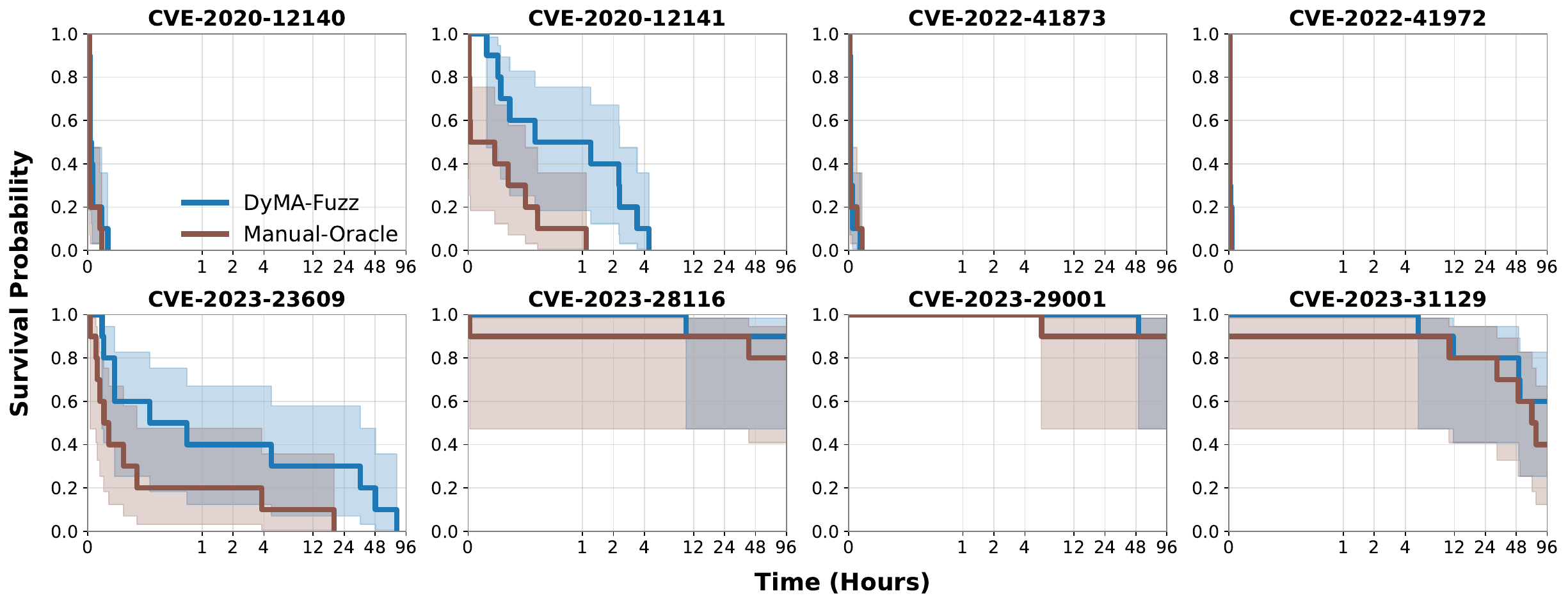}
    \caption{Bug survivability plots for previously known CVEs in Contiki-NG. The results compare \name{} with an oracle, provided manually determined information for the location of DMA input buffers.}
    \label{fig:cve_survival}
\end{figure*}

Code coverage results are shown in Table~\ref{fig:cov_tab}, with coverage-over-time shown in Figure~\ref{fig:coverage}. For binaries in the set previously used in DICE~\cite{dice2021}, similar results were seen for \name{} and DICE-MultiFuzz. 3 binaries showed either identical, or statistically insignificant difference in results. This is due to both methods accurately detecting DMA buffers and injecting data. However, there are some noteworthy cases resulting from the different methods of identifying buffer lengths. The GPS Receiver binary suffers from corruption of memory for longer inputs with MultiFuzz-DICE, discussed further in Section~\ref{sec:discussion}. The Oscilloscope and Soldering Iron observed small differences in coverage as DMA is used to sample an Analog to Digital Converter (ADC), but not every sample is read, leading to gaps in the read buffer indexes. These gaps prevent DICE from identifying the correct length of the buffer. Consequently, not all samples are fuzzed, restricting code exploration. 
The lack of detection of list-based and integrated DMA interfaces prevented DMA injection in several binaries. The Contiki and Pinetime binaries show identical coverage between DICE-MultiFuzz and MultiFuzz, while \name{} showed a significant increase in coverage.

Comparing the results of \name{} and \multifuzz{} shows the addition of DMA provides an increase in coverage across the board, statistically significant in almost every case. For binaries driven by ADC data such as the Guitar Pedal, Oscilloscope or Soldering Iron, where the input is used in calculations, but not to directly control system states, we generally see smaller improvements. Contrastingly, the GPS Receiver, Modbus, MIDI Synthesizer and Contiki binaries use structured messages of different types, with dedicated code sections to handle each message type, leading to more substantial increases in code exploration.

\semu{} was used on a subset of binaries where models were available for the MCU. Notably, for the DICE binaries, we observed low coverage, and did not observe DMA data being injected in these binaries, the causes of which are discussed further in Section~\ref{sec:discussion}. The LwIP binaries from \semu{} show an increase in coverage under \semu{} compared to \multifuzz{} operating without DMA capabilities. However, \name{} achieved similar or higher coverage compared to \semu{} in these binaries.

\subsection{DMA Input Triggered Bug Benchmarks}\label{bug_replication}
Of the 14 binaries tested in Section~\ref{coverage}, 13 have been fuzzed in prior work. Consequently, we did not uncover new bugs in these binaries, but were able to reproduce several known bugs. We consider the curated datasets of binaries with bugs triggered by DMA reported by DICE and SEmu---notably, only DICE reported bugs triggered by DMA injected data. We also consider a set of Contiki-NG versions with CVEs triggered from DMA injected data.

\begin{table}[h!]
\centering
\caption{Comparing time to find bugs discovered by \dice{}. \name{} was able to quickly and consistently reproduce bugs found by \dice{} in~\cite{dice2021}.}
\label{tab:dice_bugs}
\resizebox{\columnwidth}{!}{
\begin{tblr}{
  row{even} = {c},
  row{3} = {c},
  row{5} = {c},
  cell{1}{1} = {c=2}{},
  cell{1}{3} = {c=3}{c},
  cell{1}{6} = {c=3}{c},
  cell{3}{1} = {r},
  cell{4}{1} = {r},
  cell{5}{1} = {r},
  vline{3,6,9} = {1}{},
  vline{-} = {2-5}{},
  hline{1} = {3-8}{},
  hline{2-6} = {-}{},
}
           &                 & \textbf{\dice{}}~\cite{dice2021}        &       &       & \textbf{\name{}} (Ours)  &             &              \\
\textbf{Binary}     & \textbf{Bug Type}        & \textbf{Min}         & \textbf{Med}   & \textbf{Max}   & \textbf{Min}         & \textbf{Med}         & \textbf{Max}          \\
MIDI Synth & Free on Global  & \textbf{1m} & 4m    & 28m   & \textbf{1m} & \textbf{1m} & \textbf{1m}  \\
Modbus     & Buffer Overread & 37m         & 1h 37m & 6h 12m & \textbf{1m} & \textbf{5m} & \textbf{23m} \\
Modbus     & Buffer Overflow & 19m         & 40m   & 55m   & \textbf{1m} & \textbf{1m} & \textbf{1m}  
\end{tblr}
}
\end{table}

\noindent\textbf{DICE Bug Benchmark.~}
Given the \semu{} test dataset of binaries did not uncover bugs, we consider the reproduction of bugs and time-to-bug analysis for the set of bugs reported in DICE~\cite{dice2021}. DICE discovered bugs in two test binaries, MIDI Synthesiser and Modbus. We were able to reproduce the two cases of \textit{free()} being called with non-heap allocations in the MIDI Synthesiser binary. DICE previously reported multiple out-of-bounds accesses in the Modbus binary. \name{} was able to successfully able to replicate all crashes found in our tests with DICE. An overview of the time required to trigger these bugs in comparison to the original author's implementation of DICE is shown in Table~\ref{tab:dice_bugs}. \name{} was noticeably faster in discovering these bugs, which is expected given the higher throughput and multi-stream fuzzing mutations.

\vspace{1mm}
\noindent\textbf{DMA Input Triggered CVE-Benchmark.~}\label{cve_replication}
Prior works have relied on manual patches to replace DMA with MMIO to find bugs in real-world firmware~\cite{fuzzware2022,hoedur2023}. Several CVEs have been identified within different versions of Contiki-NG, with inputs being dependent on input from DMA. These bugs are diverse, including buffer overreads, overflows, null pointer dereferences and infinite recursion. We note prior DMA approaches either fail to identify this type of DMA, or lack published models for the required microcontroller. We test the ability for \name{} to locate the bugs found in prior MMIO works for vulnerable versions of Contiki-NG. For comparison, we create an oracle by manually identifying the DMA buffers, and marking these sections of RAM as MMIO. This oracle configuration allows MultiFuzz to mutate these DMA buffers. We monitor the environment at runtime, and immediately exit the test case if any bug other than the bug under test is detected. The tests continue until a memory corruption bug is detected, limited to 96 hours. We plot the bug survivability of our automated method, \name{}, and the oracle approach in Figure~\ref{fig:cve_survival}.

For the 8 tested CVEs, 3 could be consistently reproduced both with \name{}, and in our manually defined case within 15 minutes. These bugs have simple trigger conditions, trusting user-input indexes and sizes without validation. CVE-2020-12141 and CVE-2023-23609 were also consistently triggered by both fuzzers, but it took significantly longer to uncover these bugs due to more specific input constraints to trigger the bug, requiring processing of multiple related messages. We observe \name{'s} maximum time to trigger these bugs was approximately 4 times longer than the oracle. We expected an increase in time due to the fuzzer resets as \name{} identifies the required models, and the overhead of hooking memory. Due to the list-based structure for DMA, we also observe DMA buffers across multiple pages of memory, causing our memory hooks to be executed for a large percentage of all global variable accesses. Further investigation into the performance impacts are shown in Section~\ref{sec:overhead}.

CVE-2023-28116, CVE-2023-29001 and CVE-2023-31129 were all triggered at least once by \name{} and the manual oracle. However, neither method could trigger the CVE in all 10 trials within 96 hours of fuzzing. We note that triggering these bugs require more specific inputs, challenging to produce using random mutation. This includes: ordered sequences of operations, large specific input values, and input validity constraints. While these challenges are not attributable to DMA, the lower overhead of the manually defined oracle case did allow faster discovery of the bug in each case. CVE-2023-29001 was the bug taking longest for \name{} to discover, taking 51 hours. We find these time-frames to be within reason for real-world fuzzing.

\begin{figure}[h]
    \centering
    \includegraphics[width=\linewidth]{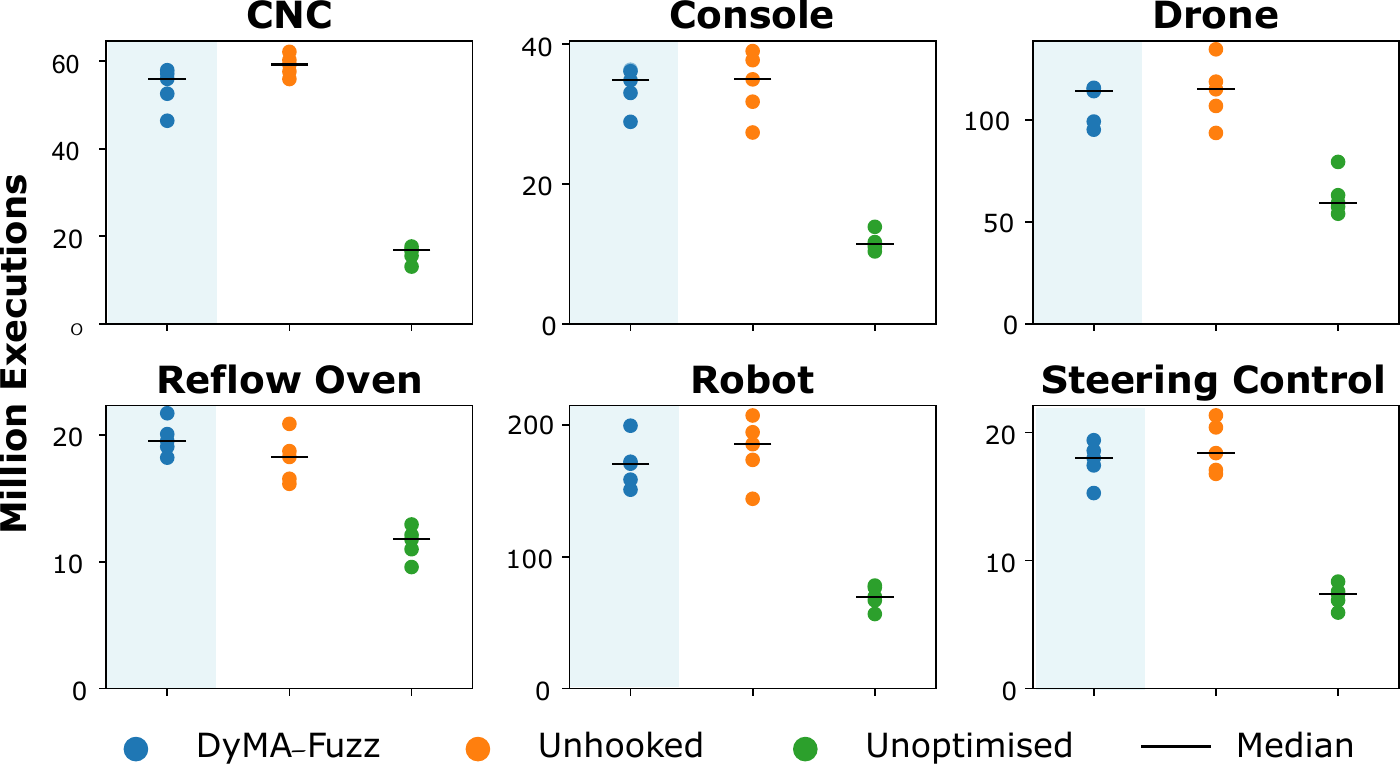}
    \caption{Execution counts across the binaries dataset from P2IM~\cite{p2im2020} without DMA. Compares \name{} with an equivalent test without DMA monitoring (Unhooked), and without dynamic growth of hooked regions (Unoptimised).}
    \label{fig:exec_overhead}
\end{figure}

\subsection{Dynamic Analysis Overhead}\label{sec:overhead}
The additional state checks performed during many loads and stores inevitably have some amount of overhead. As mentioned in Section~\ref{cve_replication}, we observed an increase in the time to trigger some bugs compared to the oracle case. In order to quantify the impact of these memory hooks, we compare throughput with DMA monitoring capabilities both enabled and disabled. We take the execution count of 5 trials, run for 12 hours on binaries without DMA. These binaries are sourced from P2IM~\cite{p2im2020}, excluding any with bugs the fuzzer is known to exploit~\cite{multifuzz2024}. Non-DMA binaries are used to ensure the same code-paths are explored for each configuration. We compare a baseline with no DMA monitoring, and compare with the execution count in the same time-frame for \name{}. We observed a negligible change in median throughput, between -8\% and +7\% across the binaries. Due to the optimisations preventing memory from being hooked unnecessarily, we expect minimal performance impact. The only memory hooked when DMA is not detected is MMIO. The results of these trials is shown in Figure~\ref{fig:exec_overhead}. We also provide an alternate result, \textit{Unoptimised}, representing the case where all RAM is hooked at all times, rather than dynamically as DMA is detected. In this unoptimised case, we observe a reduction in baseline throughput from 35\% to 72\%, depending on the binary. This represents an approximation of the overhead of \name{} compared to the manual definition of memory sections as fuzzing inputs (As done in Section~\ref{cve_replication}). We expect the overhead to scale based on the percentage of memory being monitored. However, overheads are higher in binaries that set up transfer descriptors in RAM, due to the additional monitoring required to detect linked lists of transfer descriptors. Notably, because this overhead is associated with binaries utilizing DMA, overheads for RAM hooks are limited to binaries that MMIO focused fuzzers like MultiFuzz cannot comprehensively fuzz. 

\vspace{1mm}
\begin{mdframed}
[backgroundcolor=blue!10,
linecolor=blue!60!black,
linewidth=1pt,
roundcorner=2mm] 
    \textbf{Key Takeaways}: \name{} is: i) effective at identifying DMA use across a wide range of peripherals, microcontrollers, and manufacturers, without introducing false positives; ii) is effective in uncovering real-world bugs triggered from DMA injected data without requiring manual analysis; iii) is observed to be as effective as the Manual Oracle---\textit{ideal} baseline---indicated by similar survival probabilities and profiles over 96 hours; and iii) has minimal runtime overheads due to the memory hook optimization techniques. 
\end{mdframed}

\section{Discussion}\label{sec:discussion}
\noindent\textbf{SEmu Models.~}We observed unusually low coverage from \semu{} in several of the DICE binaries even when using the published peripheral models. We observe several issues that prevent uncovering large sections of the program including missing MMIO models, missing interrupt triggers, or a lack of injected DMA data in buffers. While the issues may be resolvable by manipulating the underlying models, this requires expert knowledge of both the microcontroller peripheral domain and of the \semu{} rule system.

\vspace{1mm}
\noindent\textbf{Re-implementation of DICE.~}
In implementing DICE~\cite{dice2021} onto \multifuzz, we observed an interesting side-effect. P2IM~\cite{p2im2020}, the fuzzer DICE was built on, does not fuzz all MMIO registers. When all firmware accessed MMIO registers are moved within the scope of the fuzzer, false positives became more prevalent. Common Read-Modify-Write patterns allow the fuzzer to generate pointer-like values that can be written back to MMIO with a single bit changed, often remaining a valid pointer. While DICE's method requiring two pointer like values written to adjacent locations makes generation of false positives unlikely, across the millions of executions in a single fuzzing campaign, false positives are encountered often. Consequently, manual analysis is needed to exclude these control registers from fuzzing for DICE to work correctly on MultiFuzz in these binaries. This implies that DICE cannot be successfully applied to fuzzing frameworks that do not attempt to model register behaviour as proposed in P2IM~\cite{p2im2020}. Due to \name{'s} monitoring of non-pointer values, registers whose values are derived from fuzzer generated inputs are quickly discarded from consideration for DMA, avoiding this problem. Notably, in order to provide consistent results for comparison of coverage, any control registers that were manually categorised to assist MultiFuzz-DICE, were also applied to \name{} and \multifuzz{}. However, we validated that \name{} does not generate these false positives cases when executed without the manual models.

\multifuzz-\dice{} encountered a crash in the GPS Receiver binary due to incorrect buffer sizing. The binary uses \texttt{strlen()} to identify the length of a received string. If the fuzzer never injects a null byte, \multifuzz-\dice{} continues to grow the inferred buffer size resulting in inputs overwriting increasing regions of memory. Sufficiently long inputs cause corruption of state variables.

\vspace{1mm}
\noindent\textbf{Pointer Quirks.~}In Contiki binaries based on the CC2538 MCU, the hardware implementation specifies that pointers point to the end of the buffer, rather than the start. Due to our Ghidra analysis not having any dependence on the pointers used by the DMA controller at runtime, we still identify the buffer start locations and sizes. We provide an option to fuzz the entire buffer pointed to by the pointer, irrespective of the offset within the buffer indicated by the pointer, avoiding this issue.

\vspace{1mm}
\noindent\textbf{DMA Buffer Size Analysis.}
Ghidra, used for our static analysis of buffers, functions best with data-type and symbol information, and may require manual assistance if key information is missing. Missing type information in the LwIP binaries results in the 10 packet structures being merged into one byte array, which needs manual separation. We found removing debug symbols from the binaries had no impact on the ability for DMA buffer sizes to be determined in the Contiki binaries, and 3 of the DICE binaries. Stripping debug symbols negatively impacted 4 binaries, shrinking the inferred buffer size in each, requiring manual merging to fix. While fuzzing is still possible with reduced buffer sizes, the reachable paths may be constrained. We expect continued improvements to reverse engineering tools will further improve handling of stripped and closed-source binaries. Pinetime uses stack buffers for DMA transfers, predominantly 1 or 4 bytes. Our fallback method caught these cases. We discuss the possibility of identifying the sizes of stack buffers, and the consequences in Section~\ref{sec:future_work}.

\subsection{Threats to Validity}
\name{} operates under the assumption that all usage of DMA is for input/output with peripherals. While this is exceptionally common, DMA access can be performed for other purposes. For example, a hardware replacement for \texttt{memcpy()}, which may be used to copy data from ROM or other locations in RAM. In these cases, the destination buffer will be populated with fuzzing data, rather than values from the source location. We did not observe memory-to-memory DMA operations in the tested real-world binaries, and 13 of the 14 binaries contain \texttt{memcpy()} which uses the CPU for copy operations. The DMA initialization and synchronization processes make DMA impractical for small, synchronous transfers typically done using \texttt{memcpy()}.

Another possibility is the use of peripherals that monitor the memory bus, without performing reads or writes. Peripherals such as a memory watch unit monitor the memory bus and generate events when certain regions are accessed. \name{} would erroneously mark these regions as DMA inputs and mutate them.

As detailed in Section~\ref{identify_registers}, we rely on identifying which values written to MMIO are pointers, and which are invalid pointers. For microcontrollers with narrower address spaces, this becomes more challenging. On a 16-bit MCU, a much larger proportion of word-sized value-space is covered by the valid RAM address-space, increasing the risk of false positives. We note that prior works~\cite{dice2021} found that 8 or 16 bit MCUs typically do not feature DMA.

We assume that DMA is implemented correctly by firmware developers. Our approach attempts to detect DMA transfers, but does not validate that the configuration is suitable for the target hardware. Buffer overflow bugs based on setting a DMA transfer count greater than the buffer size would not be caught by \name{}. We consider this a minor limitation, as manufacturers typically provide robust implementations with clear interfaces as part of a Hardware Abstraction Layer (HAL). Therefore, bugs in data processing are likely to be far more prevalent.

\section{Related Work}
One avenue of research considers employing the target or similar platforms~\cite{avatar2014,avatar2018,conware2021,pretender2019,gdbfuzz2023,uafl2022,shift2024,inception2018} for fuzzing monolithic firmware. Largely for performance and binary-only reasons, many recent efforts delve into solving fuzzing challenges in re-hosting environments~\cite{halucinator2020,p2im2020,fuzzware2022,emberio2023,safirefuzz2023,splits2023,uemu2021,hoedur2023,multifuzz2024,laelaps2020,jetset2021,icicle2023,splits2023,basesafe2020,pararehosting2021}. Our work tackles the problems faced in re-hosting environments when the target firmware employs DMA for peripheral interactions.

\vspace{1mm}
\noindent\textbf{MMIO Data Injection.~}Existing work focusing on the injection of data at architecture defined hardware interfaces such as Memory Mapped Input/Output (MMIO) and interrupts has been explored from many angles, with the aim of minimising unnecessary or invalid inputs~\cite{p2im2020,uemu2021,emberio2023,fuzzware2022,hoedur2023,multifuzz2024}. Current state-of-the-art methods for injecting data into MMIO employ a stream-based input representation to better align the input generation with how fuzz inputs are consumed by the firmware~\cite{hoedur2023,multifuzz2024}, rather than a single file~\cite{afl2010,aflplusplus2020}. However, the scope of these works does not extend to tackling the complexities of emulating DMA inputs.

\vspace{1mm}
\noindent\textbf{DMA Data Injection.~}To address the problem with complex interfaces in DMA, DICE~\cite{dice2021} extends P2IM~\cite{p2im2020} to detect type $I_M$ DMA transfer descriptors based on automated runtime analysis. Once these transfer descriptors are detected, DICE injects fuzzer data, assuming reads to adjacent bytes of consistent size are performed from the location found in the transfer descriptor and cover the entire target buffer. The study in~\cite{perry2024} uses an existing HAL implementation to infer hardware behaviour, and develop a model for hardware interactions in the re-hosted environment. SEmu~\cite{semu2022} similarly uses external information from manufacturer datasheets to define hardware models, and specifically targets fuzzing. Natural language processing is used to generate a rule-set to define the behaviour of each peripheral. Inferring models reduces the amount of manual effort to provide high-fidelity peripheral emulation, but often requires manual assistance to correctly function with DMA. We also acknowledge GDMA~\cite{gdma2025}, a concurrent work modeling DMA using access patterns.

\section{Conclusion and Future Work}\label{sec:future_work}

Future work may wish to extend the scope of identifying buffers to include the heap and stack. Runtime monitoring of allocators and stack frames could provide a more robust basis for size inference than our fallback method based on sequential access. While heap allocators could be identified using HEAPSTER~\cite{heapster2022}, and stack allocations extracted from Ghidra~\cite{ghidra}, the additional tracing of current allocations would increase overheads, and we find most binaries do not perform DMA transfers to non-global buffers.

We found \name{}, a DMA enabled fuzzing framework based on runtime analysis of RAM and MMIO accesses to be an effective tool for monolithic firmware testing. It succeeded in achieving high code coverage across a wide range of binaries, using DMA interfaces in MMIO, RAM and integrated into other peripherals. We were able to reproduce known bugs and CVEs in a timely manner, demonstrating real-world suitability for fuzzing.

\begin{acks}
The work has been supported by the Australian Government’s Research Training Program Scholarship (RTPS). The authors would also like to thank the anonymous reviewers for their valuable comments and helpful suggestions.
\end{acks}

\bibliographystyle{ACM-Reference-Format}
\bibliography{references}

\end{document}